\documentclass[pra,twocolumn,showpacs,floatfix]{revtex4-1}
\usepackage{natbib,times}
\usepackage{graphicx}
\usepackage{dcolumn}
\usepackage{amsmath}
\usepackage{amssymb}
\usepackage{color}

\begin{document}

\title{Dissipative dynamics of quantum correlations in the strong-coupling regime}
\author{Ferdi Altintas}\email{ferdialtintas@ibu.edu.tr}  \author{Resul Eryigit}\email{resul@ibu.edu.tr}
\affiliation{Department of Physics, Abant Izzet Baysal University, Bolu, 14280, Turkey.}
\begin{abstract}
The dynamics of entanglement and quantum discord between two identical qubits strongly interacting with a common single mode leaky cavity field have been investigated beyond the rotating wave approximation (RWA) by using recently derived Lindblad type quantum optical master equation [F.~Beaudoin, J.M.~Gambetta, A.~Blais, Phys. Rev. A {\bf 84}, 043832 (2011)] that can describe the losses of the cavity field in a strong atom-field coupling regime. Contrary to previous investigations of the same model in the dissipative regime by using the 
standard Lindblad quantum optical master equation in a strong-coupling regime, the atom-field steady states are found to be cavity decay rate independent and have a very simple structure determined solely by the overlap of initial atomic state with the subradiant state which is valid for all coupling regimes. Non-RWA dynamics are found to have remarkable effects on the steady state quantum discord and entanglement that cannot be achieved under RWA conditions, for instance, they can induce steady state entanglement even for the initial states that have no overlap with the subradiant state. Moreover, the non-RWA dynamics are found to reverse the initial state dependence of steady state entanglement and quantum discord contrary to the RWA case.
\end{abstract}

\maketitle
\section{Introduction}
The description of light-matter interaction is the fundamental question investigated in circuit and cavity quantum electrodynamics~(QED). The single-mode spin-boson model, the so-called quantum Rabi model, is the simplest possible physical model that describes the interaction of a two-level atom~(a qubit) with a quantized single mode electromagnetic field~(a harmonic oscillator)~\cite{rabi}. Indeed, this simple model has been used to understand a wide variety of phenomena in quantum optics and condensed matter systems, such as quantum dots, trapped ions, superconducting~(artificial) qubits, optical and microwave cavity QED, among others~\cite{ashhab,rbrev,rbused}. Despite its old age and numerous investigations over the last few decades, the spectrum of the Rabi model has been given only recently in a closed form as the solution of a transcendental equation with a single variable~\cite{rbsol}. On the other hand, the Rabi model is quite simplified and can be solved analytically when the excitation non-conserving terms, the so-called counter-rotating terms that are responsible for simultaneously exciting or de-exciting the atom and field, are ignored~\cite{jcmodel}. This approximation is known as the rotating wave approximation~(RWA) which reduces the Rabi model to the Jaynes-Cummings model and has been widely used in quantum optical settings. The RWA is valid only at weak atom-field coupling strengths and nearly resonant conditions. Indeed, the typical quantum optics experiments are performed at microwave to visible light range of electromagnetic wave spectrum where the atom-field coupling strength is several orders of magnitude smaller than the transition frequencies of the atom and the field, so that RWA is inherently justified for those parameters. Despite being an approximation, the analytical achievements of RWA have successfully explained many novel phenomena and experimental results, such as Rabi oscillations~\cite{burne}, squeezing~\cite{jrkjlm}, non-classical~\cite{burne2} and Fock~\cite{wvhw} states, collapse and revivals of atomic inversion~\cite{grhw}, and entanglement between atoms as well as atom and field systems~\cite{sppk}.  However, with the recent developments in the area of circuit and cavity QED systems~\cite{rbrev,rbused}, ultra and deep strong light-matter couplings became experimentally achievable which makes it necessary to use full the Rabi model to explain all the dynamical and statistical properties of these systems.

The discussion of the role of counter-rotating terms on the atomic dynamics has received  plenty of attention, recently~\cite{wddb,hnkb,phoenix,lemb,bloch,mag}. Although the effects of these terms are extremely small and averages to zero over a very short time scale for the weak coupling regime, they, nonetheless, cause significant shifts in the dynamics of the system at intermediate to deep interaction strength region. Generation of photons~\cite{wddb} and entanglement~\cite{hnkb} from zero excitation initial state, bifurcation in the phase space~\cite{phoenix}, a fine structure in the optical Stern-Gerlach effect~\cite{lemb}, Bloch-Siegert shift~\cite{bloch} and chaos in the Dicke model~\cite{mag} are the remarkable examples of these novel effects and none of them are observed under RWA. On the other hand, the significance of the counter-rotating terms, especially for multipartite and dissipative systems, is still under debate and needs further investigation.

Any realistic quantum system inevitably interacts with the outside world which leads to decoherence. In circuit and cavity QED systems with atom-field interactions, this unavoidable interaction leads to qubit relaxation and dephasing as well as cavity relaxation. The standard Lindblad quantum optical master equation~\cite{lindbadme} is the general form of a Markovian master equation that can describe the irreversible and non-unitary processes of the atom-field system~\{see Eq.~(14) in Ref.~\cite{bgb}\}. While deriving the standard dissipators in this master equation, the qubit-field interaction is neglected and therefore separate interactions of the involved parts with the environment are taken into account. For the weak atom-field coupling regime where RWA can be safely applied, this master equation works well and can be used  to predict the experimental results in cavity and circuit QED systems~\cite{slmeexp}. On the other hand, it is controversial to use the standard Lindblad master equation without any formal proof to describe the dephasing and relaxation processes in strong coupling regime where RWA is no longer valid. In fact, qubit-field coupling becomes influential and leads to unphysical effects~\cite{bgb}; even at zero temperature, the standard master equation with the Rabi model produces spurious qubit flipping and creates excess photons in the atom-field system. It was shown in Ref.~\cite{bgb} that it is possible to correct the predictions of the Rabi model in the dissipative case by carefully deriving a new Lindblad type dissipator which involves transitions between the eigenstates of the Rabi model. The improved Lindblad master equation given in Ref.~\cite{bgb} is expected to be a new avenue in the studies of strong light-matter interactions including dissipative and multipartite systems and to help understanding the role of counter-rotating terms under dissipation and dephasing processes.

In the present study, we extend the above ideas by studying a model of two cold-trapped atoms resonantly interacting with a single-mode leaky cavity field. This model is quite versatile and can be directly verifiable in circuit and cavity QED setups. Under RWA, the dynamics of the atom-field system under losses have been investigated thoroughly and the atom-field steady states have a very interesting structure determined solely by the overlap between the initial state and the decoherence free state~\cite{sfzgp,mspsg,zfrt,fcah}. On the other hand, the same model has also been investigated under non-RWA conditions by using the standard Lindblad master equation~\cite{fjl,jlmz,jlf,fare}. Since the master equation with standard dissipators cannot explain properly the effect of dissipation in the strong coupling regime, the reported role of the counter-rotating terms should be reconsidered. By using the improved Lindblad master equation given in Ref.~\cite{bgb} and the structure of atom-field steady states under RWA, we provide a clear picture for the role of counter-rotating terms on atom-field steady states and also give their analytical expressions for the general class of two-qubit initial states, the so-called "extended" Werner-like states~\cite{bellomo08} which may reduce to the Werner-like mixed states or to the Bell-like pure states for certain conditions. We also reexamined some of the main results obtained by the standard Lindblad master equation in the strong atom-field coupling regime and revised them by using the results obtained by the improved Lindblad master equation. Beside this, we also analyzed the role of counter-rotating terms on the dynamics of quantum correlations, such as entanglement and quantum discord between two atoms and found that the counter-rotating terms can, peculiarly, reverse the initial state dependence of the steady state entanglement and quantum discord contrary to the RWA case. For RWA dynamics, the overlap of initial state with the decoherence free state is crucial to obtain a long-lived quantum correlated state under losses~\cite{sfzgp,mspsg,zfrt,fcah}. We showed that the counter-rotating term contributions break down this description and lead to long-lived quantum correlated state without this overlap. 

The paper is structured as follow. In Sec.~\ref{shme}, we present the system Hamiltonians for both RWA and non-RWA conditions and the master equations that describe the losses of the cavity field. The correlation measures, quantum discord and entanglement, are also briefly discussed in this section. In Sec.~\ref{rslt}, we present our results. We conclude with a brief summary of important results in Sec.~\ref{conc}.

\section{Hamiltonian and the master equation}\label{shme}
The Rabi model describes the interaction of a two-level atom with a  single mode electromagnetic field. The Hamiltonian can be given as~(with $\hbar=1$ throughout)~\cite{rabi}:
\begin{eqnarray}\label{rh}
H_R=\frac{\omega_a}{2}\sigma_z+\omega_ra^{\dagger}a+g(\sigma_++\sigma_-)(a+a^{\dagger}),
\end{eqnarray}
where $\omega_a~(\omega_r)$ is the transition frequency for the atom~(cavity field), $\sigma_z$ is the qubit energy operator, $\sigma_{\pm}$ are the spin-flip operators and $g$ is the atom-field coupling constant.

The terms $a^{\dagger}\sigma_+$ and $a\sigma_-$ that appear in the interaction part of the Rabi Hamiltonian  are called counter-rotating terms and they describe "virtual processes"; $a\sigma_-$ describes a process in which a photon is annihilated in the cavity mode as the atom makes a transition from its excited to the ground state, while $a^{\dagger}\sigma_+$ describes the creation of a photon in the cavity mode as the atom makes an upward transition. So, the counter-rotating terms do not conserve the total number of excitations in the system. 

The virtual processes have observable effects when $g$ is comparable with $\omega_a$ and $\omega_r$, i.e., in strong atom-field coupling regime, however when the weak coupling, $g<<\omega_a,\omega_r$, and the nearly resonant, $|\Delta|=|\omega_a-\omega_r|<<\omega_a,\omega_r$, conditions are simultaneously satisfied, these effects become negligible and the counter-rotating terms can be neglected in the interaction part of the Rabi Hamiltonian. This approximation is known as the rotating wave approximation~(RWA) and the Rabi model reduces to the well known Jaynes-Cummings model given by the Hamiltonian~\cite{jcmodel}:
\begin{eqnarray}\label{jch}
H_{JC}=\frac{\omega_a}{2}\sigma_z+\omega_ra^{\dagger}a+g(\sigma_+a+\sigma_-a^{\dagger}).
\end{eqnarray}
In opposition to the Rabi Hamiltonian, the total number of excitations in $H_{JC}$ is a conserved quantity which makes $H_{JC}$ analytically accessible.

In cavity QED systems, there would be photon loss from the cavity due to the imperfections in the cavity mirrors. At absolute zero temperature~($T=0$) and under Markov approximation, this can be represented by the standard Lindblad quantum optical master equation~\cite{lindbadme}:
\begin{eqnarray}\label{slme}
\dot{\rho}=-i[H,\rho]+\kappa D[a]\rho,
\end{eqnarray}
where $D[m]\rho=\frac{1}{2}\left(2m\rho m^{\dagger}-\rho m^{\dagger}m-m^{\dagger}m\rho\right)$, $\kappa$ is the photon leakage rate, $m$ is the appropriate dissipator operator, and $\rho$ is the atom-field density matrix. The first term in Eq.~(\ref{slme}) describes the unitary Hamiltonian dynamics, while the second part accounts for the dissipation. At the weak-coupling regime, where RWA can be applied, the dissipator part of the master equation is well known and involves field annihilation operator, $a$, which is responsible for the leakage of photons from the cavity. Indeed, at weak-coupling regime where RWA can be applied, this master equation can be safely used and the dissipation can bring the system to the ground state, $\left|g0\right\rangle$, of $H_{JC}$.

Although it has been widely investigated by many groups, recently~\cite{wddb,fjl,jlmz,jlf,fare,seke1,seke2,seke3,brcpls}, using the above dissipator in the dissipative Rabi model in an {\it ad hoc} manner has been shown to lead to unphysical consequences~\cite{bgb}; at $T=0$, the dissipation will generate excess excitations when even no energy is added to the system and consequently the dissipation will drive the system out of the ground state of $H_R$. In fact, the main failure of this master equation in the strong-coupling regime is to neglect the qubit-field coupling when deriving the dissipator part of the master equation. In Ref.~\cite{bgb}, the qubit-field coupling have been included in the derivation of the master equation and the new Lindblad master equation that describes the losses of the cavity field at $T=0$ and strong coupling regime under Markov approximation can be given as:
\begin{eqnarray}\label{newme}
\dot{\rho}=-i[H_R,\rho]+\displaystyle\sum_{j,k>j}\Gamma_{\kappa}^{jk}D\left[\left|j\right\rangle\left\langle k\right|\right]\rho,
\end{eqnarray}
where $\Gamma_{\kappa}^{jk}$ are the relaxation coefficients and are related to the spectral density of the bath and the system-bath coupling constant~\cite{bgb}. It can be simplified as~\cite{rlsh}, $\Gamma_{\kappa}^{jk}=\kappa\frac{\omega_k-\omega_j}{\omega_r}|\left\langle j\right|(a+a^{\dagger})\left|k\right\rangle |^2$. Here $\left|j\right\rangle$ and $\omega_j$ are the eigenstates~(dressed basis) and the eigenvalues of the Rabi Hamiltonian, respectively, i.e., $H_R\left|j\right\rangle=\omega_j\left|j\right\rangle$ and the eigenstates are labeled according to increasing energy, i.e., label $\left|j\right\rangle$ such that $\omega_k>\omega_j$ for $k>j$. Indeed, as shown in Ref.~\cite{bgb}, the new dissipator drives the atom-field system to the true ground state. Also, note that the standard dissipator of Eq.~(\ref{slme}) can be obtained from that of Eq.~(\ref{newme}) in the limit $g\rightarrow 0$.

Since the strong atom-field coupling regime became relevant with the development of the recent technology~\cite{rbrev,rbused} and a number of investigations of the dissipative Rabi model with standard dissipator have been reported in the literature in the strong-coupling regime~\cite{fjl,jlmz,jlf,fare,seke1,seke2,seke3,brcpls}, a reexamination of these studies in the light of the improved master equation~(\ref{newme}) is due. Toward that goal, in the present study, we will analyze the dynamics of entanglement and quantum discord between two identical qubits~(A and B) interacting with a common single mode leaky cavity field. The Hamiltonians for non-RWA and RWA cases can be found by just replacing $\sigma_z\rightarrow\sigma_z^A+\sigma_z^B$ and $\sigma_{\pm}\rightarrow\sigma_{\pm}^A+\sigma_{\pm}^B$ in Eqs.~(\ref{rh}) and~(\ref{jch}), respectively. We will solve the master equations~(\ref{slme}) and~(\ref{newme}) numerically for the considered Hamiltonians and the initial states where the atoms are one of the four extended Werner-like states~\cite{bellomo08} and the cavity field is in its vacuum:
\begin{eqnarray}\label{initialstate}
\rho_{\Psi_{\alpha}^{\pm}}(0)&=&\left[\frac{1-r}{4}I_{AB}+r\left|\Psi_{\alpha}^{\pm}\right\rangle\left\langle \Psi_{\alpha}^{\pm}\right|\right]\otimes\left|0\right\rangle\left\langle 0\right|,\nonumber\\
\rho_{\Phi_{\alpha}^{\pm}}(0)&=&\left[\frac{1-r}{4}I_{AB}+r\left|\Phi_{\alpha}^{\pm}\right\rangle\left\langle \Phi_{\alpha}^{\pm}\right|\right]\otimes\left|0\right\rangle\left\langle 0\right|,
\end{eqnarray}
where $I_{AB}$ is the $4\times 4$ identity, $r~(0\leq r\leq 1)$ is the purity of the initial states,  
$\left|\Phi_{\alpha}^{\pm}\right\rangle=\alpha\left|eg\right\rangle\pm\sqrt{1-\alpha^2}\left|ge\right\rangle$ and $\left|\Psi_{\alpha}^{\pm}\right\rangle=\alpha\left|ee\right\rangle\pm\sqrt{1-\alpha^2}\left|gg\right\rangle$ are the Bell-like states, and $\alpha~(0\leq\alpha \leq 1)$ is called the degree of correlations. The extended Werner-like states are mixed and reduce to the well known Werner-like mixed states at $\alpha=1/\sqrt{2}$ and to the Bell-like pure states at $r=1$. The atomic reduced density matrix can be obtained by taking a partial trace of atom-field density matrix over the cavity degrees of freedom. The above initial states in the standard basis $\{\left|1\right\rangle\equiv\left|ee\right\rangle,\left|2\right\rangle\equiv\left|eg\right\rangle, \left|3\right\rangle\equiv\left|ge\right\rangle, \left|4\right\rangle\equiv\left|gg\right\rangle\}$ have an X structure in the atomic space with non-zero elements only in its main- and anti-diagonals:
\begin{eqnarray}
\label{xmatrix}
\rho_{AB}=\left (\begin{array}{cccc} \rho_{11}  & 0 & 0  & \rho_{14} \\ 0  & \rho_{22} & \rho_{23}  & 0 \\ 0  & \rho_{32} & \rho_{33}  & 0 \\ \rho_{41}  & 0 & 0  & \rho_{44} \end{array} \right) \ .
\end{eqnarray}
It was shown that the standard Lindblad master equation~(\ref{slme}) for both RWA and non-RWA Hamiltonians preserves the X structure of the density matrix~\cite{fjl,jlf,fare}. We have checked that the X structure of the atomic reduced density matrix remains intact also under the improved master equation~(\ref{newme}).

Since the excitation number in the system is not conserved for the Rabi model, it is worth to explain briefly the numerical technique to solve the master equations~(\ref{slme}) and~(\ref{newme}) for the two-qubit Rabi Hamiltonian. In the present study, we have carefully checked the convergence of computed properties versus the dimension of the cavity Fock field. We have considered basis vectors of type $\left|i,j,n\right\rangle$, where $i,j=e,g$ and $n=0,1,2,\ldots M$ to obtain a set of differential equations for the density matrix elements. It is found that $M\approx 50$ gives well-converged results for the highest interaction strength considered in the present work. Moreover, depending on the interaction strength, 100-150 dressed basis of the two-qubit Rabi Hamiltonian are used to calculate the time evolution. During the numerical calculations, the basic properties of density matrix, such as positivity, hermiticity, and trace preservation properties are all monitored.

Our next step is to determine the correlations between the atoms. The entanglement, a kind of quantum correlation, determines whether or not a given bipartite state is separable. Entangled states are broadly accepted as a necessary resource for a set of quantum tasks in quantum information theory~\cite{entuse1,entuse2}, such as quantum key distribution and teleportation. Entanglement can be calculated through entanglement of formation~(EoF) which is the function of the entanglement monotone, the so-called concurrence, for bipartite states as~\cite{wootters}:
\begin{eqnarray}\label{eof}
\mathcal{C}_{EoF}(\rho_{AB})=-\eta\log_2\eta-(1-\eta)\log_2(1-\eta),
\end{eqnarray}
where $\eta=1/2\left(1+\sqrt{1-C^2}\right)$ with $C=2\max\{0, |\rho_{14}|-\sqrt{\rho_{22}\rho_{33}}, |\rho_{23}|-\sqrt{\rho_{11}\rho_{44}}\}$ being the concurrence for X states. On the other hand, quantum discord~(QD) emerged as a new fundamental type of quantum correlation beyond entanglement~\cite{howz,lhvv,mbcpv}.  QD as a fundamental resource is shown to be useful in practical quantum information tasks, such as DQC1~(deterministic quantum computation with one quantum bit)~\cite{dqc1}, Grover search~\cite{grover} algorithms and remote state preparation~\cite{rsp}. Its definition is based on the difference between quantum versions of two classically equivalent definitions of mutual information. Non-zero QD signifies that it is impossible to extract all information about one subsystem by performing a set of measurements on the other subsystem. QD can also be calculated, analytically, for X states as~\cite{wlnl}:
\begin{eqnarray}\label{qd}
\mathcal{C}_{QD}(\rho_{AB})=\min\{Q_1,Q_2\},
\end{eqnarray}
where  $Q_j=h[\rho_{11}+\rho_{33}]+\sum_{k=1}^4\lambda_k\log_2\lambda_k+D_j$
with $\lambda_k$ being the eigenvalues of $\rho_{AB}$ and $h[x]=-x\log_2x-(1-x)\log_2(1-x)$ is the binary entropy.
Here $D_1=h[\tau]$, where $\tau=\left(1+\sqrt{[1-2(\rho_{33}+\rho_{44})]^2+4(|\rho_{14}|+|\rho_{23}|)^2}\right)/2$ 
and $D_2=-\sum_{k=1}^4 \rho_{kk}\log_2\rho_{kk}-h[\rho_{11}+\rho_{33}]$. QD and EoF are equal for pure states, but the relation is complicated for the mixed states; there are separable mixed states with non-zero QD~\cite{facca}. Recently, the comparison attempts for the dynamics of quantum discord and entanglement have received a great deal of attention for open quantum systems~\cite{gds1,gds2,gds3,gds4,gds5,gds6,gds7,gds8,gds9,gds10,gds11}. The studies show that QD is  much more robust compared to entanglement under dissipative and dephasing processes where entanglement can suffer sudden death.

In the following, our main concerns will be threefold. Since  numerous studies have been recently based on the standard master equation~(\ref{slme}) in the strong-coupling regime~\cite{fjl,jlmz,jlf,fare,seke1,seke2,seke3,brcpls}, our first thought is to reexamine some of the main results obtained in these studies by using the new Lindblad master equation~(\ref{newme}). Since the standard dissipator cannot describe the losses in the strong atom-field coupling regime, the role of counter-rotating terms on the dissipative dynamics of the atom-field system are far from being understood. As the second issue, we will provide a comprehensive picture for the atom-field steady states and give their analytical expressions for the extended Werner-like initial states. Our final focus is to analyze the role of counter-rotating terms on the dynamics of entanglement and quantum discord. In the following, we will restrict ourselves to the resonant case, $\omega_a=\omega_r=\omega$.
\section{Results}\label{rslt}
\begin{figure*}
  \begin{center}
	{\scalebox{0.7}{\includegraphics{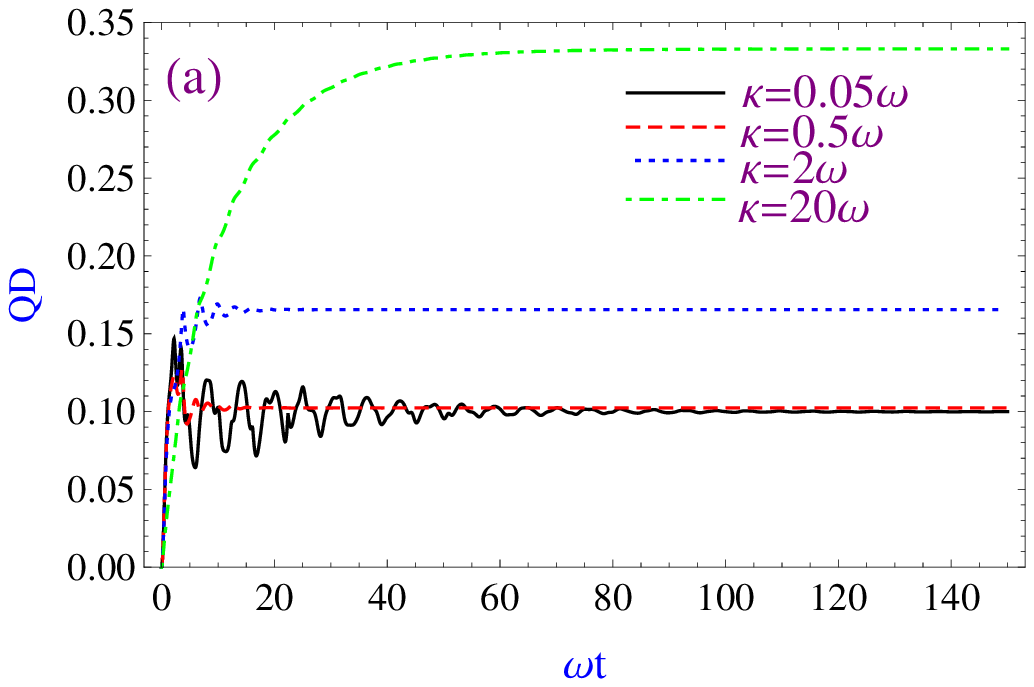}}}
  {\scalebox{0.7}{\includegraphics{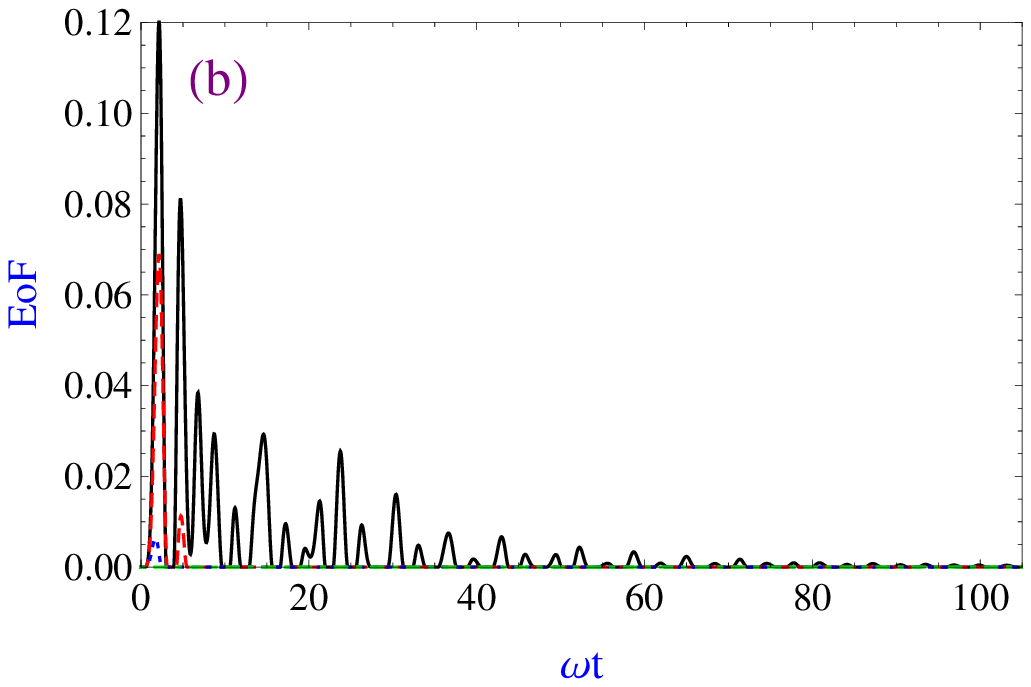}}}

 {\scalebox{0.7}{\includegraphics{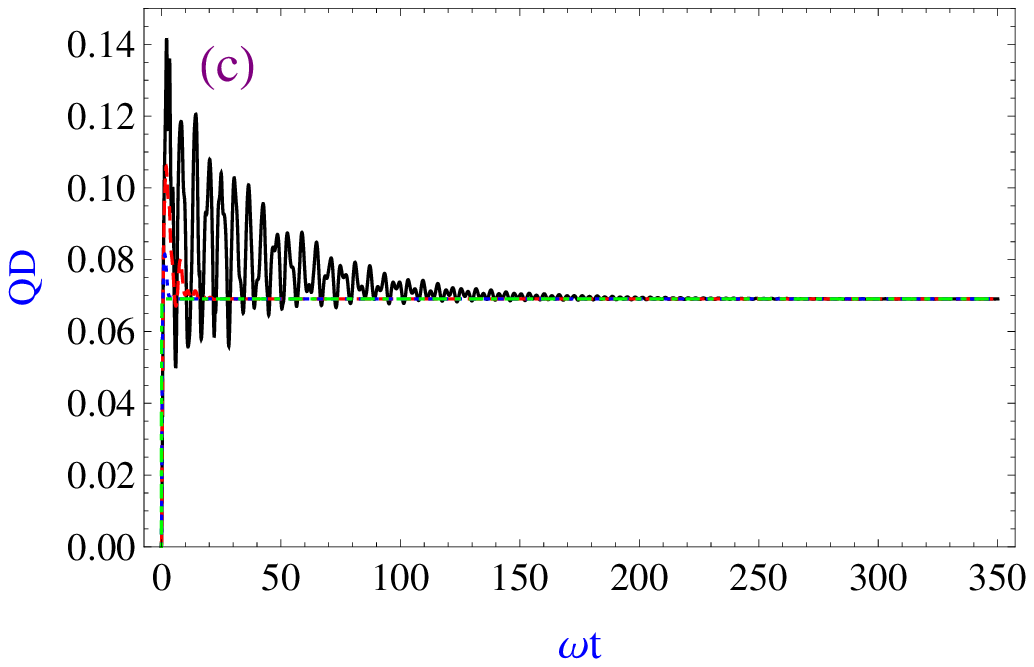}}}
 {\scalebox{0.7}{\includegraphics{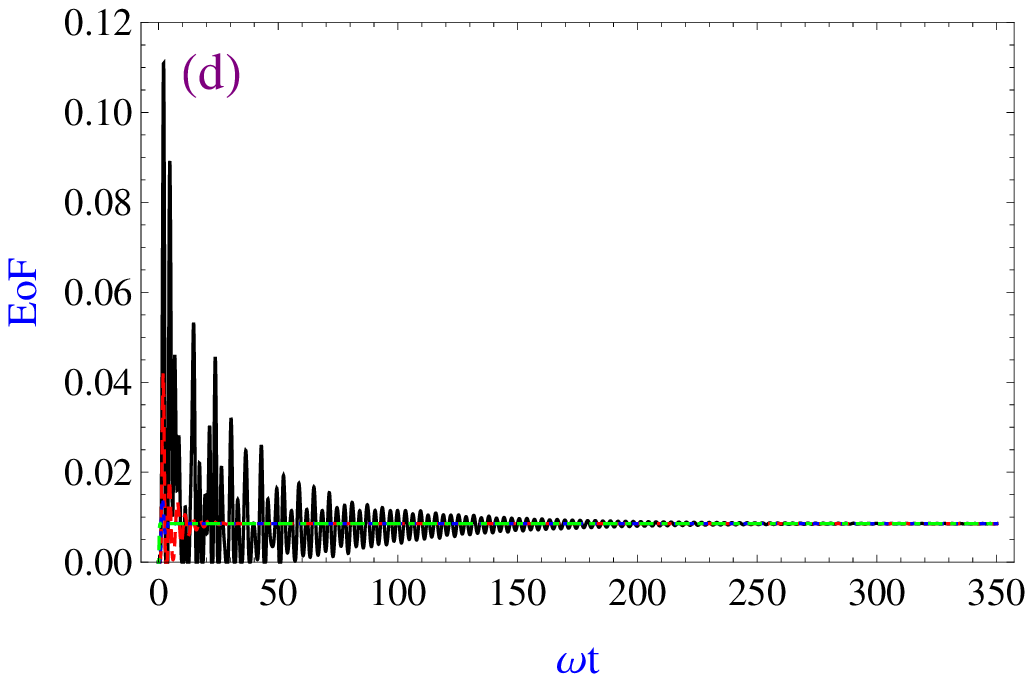}}}
  \end{center}
\caption{(Color online) The effect of cavity decay rate $\kappa$ (a) and (c) on quantum discord and (b) and (d) entanglement~ versus $\omega t$ for $\left|gg0\right\rangle$ initial state, $\kappa=0.05\omega$~(black, solid), $\kappa=0.5\omega$~(red, dashed), $\kappa=2\omega$~(blue, dotted), and $\kappa=20\omega$~(green, dot-dashed) in strong coupling regime with $g=0.35\omega$. The top figures are obtained by using the standard dissipator Eq.~(\ref{slme}), while the bottom plots are for the improved dissipator~Eq.~(\ref{newme}).} 
\end{figure*}

We start our qualitative analysis by examining the main disagreements between the two master equations in the strong-coupling regime. In Fig.~1, we display the effect of cavity decay rate, $\kappa$, on the dynamics of QD and EoF for the vacuum~$\left(\left|gg0\right\rangle\right)$ initial state at $g=0.35\omega$. The results in the top plots are obtained by the use of the standard master equation~[Eq.~(\ref{slme})], while the bottom plots display the results for the new dissipator~[Eq.~(\ref{newme})] in the case of two-qubit Rabi Hamiltonian. Since the vacuum state contains no initial excitations to start with and the Jaynes-Cummings Hamiltonian conserves the total excitation number in the atom-field system, no type of correlations will be induced in the time evolution under RWA. On the other hand, the counter-rotating terms in the Rabi Hamiltonian will generate virtual excitations and consequently non-zero correlations between the atoms which can be trapped at a non-zero value under cavity losses for both cases. If we study the effect of cavity losses on the dynamics of correlations by using the standard dissipator~[Figs.~1(a) and~1(b)] as was done in Refs.~\cite{fare,seke1,seke2,seke3}, the results seem to be quite surprising; as $\kappa$ increases, although the disentanglement time and the maximum of EoF decreases, the steady state QD, peculiarly, increases with $\kappa$ and is saturated in the maximum possible value~($\mathcal{C}_{QD_{max}}=1/3$) in the absence of entanglement for a very bad quality cavity case~($\kappa/\omega>20$)~\cite{fare}. Although these results seem to indicate a surprising possibility of dissipation enhancement of quantum discord~\cite{fare}~(as well as atomic energy as found in Refs.~\cite{seke1,seke2,seke3}) to its maximally possible value for a separable state, its source is the error in the dissipator which creates excess excitations in the steady state. Considering the role of $\kappa$ on the dynamics of QD~[Fig.~1(c)] and EoF~[Fig.~1(d)] for the new dissipator demonstrates that the steady states are $\kappa$ independent; QD and EoF reaches 0.069 and 0.0086 in the long time limit, respectively, regardless of the magnitude of $\kappa$. In fact, $\kappa$ only determines how fast the correlations reach their steady state values. This is quite understandable, since $\kappa$ is just a parameter that determines how fast the photons can escape from the cavity mirrors, it should control the speed of reaching the steady state rather than its magnitude in the long time limit. Bath decay dependent steady state shifts of several properties of the system in the strong-coupling regime have been recently investigated by using the standard dissipator~\cite{fare,seke1,seke2,seke3,brcpls}. Moreover, Ma {\it et al.} in Ref.~\cite{mswn} studied the dynamics of entanglement between two qubits strongly interacting with a common Lorentz-broadened cavity mode at zero temperature by using a more general master equation approach, the so-called hierarchical equation method without employing rotating wave, Born, and Markovian approximations. One should note that the approach in Ref.~\cite{mswn} is quite different compared to the standard Lindblad master equation case, but still the asymptotic values of entanglement are found to be dissipation strength dependent, which is in contradiction to our findings. The simple example shown in Fig.~1 indicates that some of the previously obtained results with the Rabi model need to be reexamined. As shown in Fig.~1(b) for the standard dissipator case, the atomic steady states become separable as $\kappa>0.05\omega$~\cite{fare}, while the steady states can posses entanglement for the new dissipator case~[Fig.~1(d)].

Now, we will consider the steady states of RWA and non-RWA dynamics. The steady state density matrix under cavity losses and RWA dynamics investigated by the standard master equation~(\ref{slme}) are known to be $\kappa$-independent~\cite{sfzgp,mspsg}. As shown in the bottom plots of Fig.~1, they are also $\kappa$-independent in the case of non-RWA dynamics with  the improved dissipator. Therefore, in the following, we will set $\kappa=0.1\omega$.

The dynamics under RWA with losses given by Eq.~(\ref{slme}) have been investigated by many groups, which indicate that the atom-field steady states have a simple structure. We follow the ideas developed in Refs.~\cite{sfzgp,mspsg,zfrt,fcah} to determine those states. The cavity decay tends to throw out any initial excitations from the cavity mirror and consequently it drives the atom-field system to the ground state~$\left(\left|gg0\right\rangle\right)$ of two-qubit $H_{JC}$. On the other hand, there exist a highly entangled subradiant state~(decoherence-free state) for this model due to the interaction of the atoms with a common environment which leads to an indirect qubit-qubit interaction~\cite{sfzgp,mspsg}. This subradiant state for this model can be simply determined by finding the eigenvector of two-qubit $H_{JC}$ for zero eigenvalue, which is the maximally entangled~(Bell) state in the atomic subsystem given as $\left|\Phi^-0\right\rangle=\frac{1}{\sqrt{2}}\left(\left|eg0\right\rangle-\left|ge0\right\rangle\right)$. If an initial atomic state has a non-vanishing overlap with the subradiant state, $b=\left\langle \Phi^-\right|\rho_{AB}(0)\left|\Phi^-\right\rangle$, the amount of overlap determined by $b$ would be trapped in the subradiant state, while the remaining part, $1-b$, would decay to the ground state. Therefore, depending on the specific initial state~\cite{expss}, the atom-field steady states would be a combination of $\left|gg0\right\rangle$ and $\left|\Phi^-0\right\rangle$ which can be written as:
\begin{eqnarray}\label{ssgf}
\rho^{SS}=(1-b)\left|gg0\right\rangle\left\langle gg0\right|+b\left|\Phi^-0\right\rangle\left\langle\Phi^-0\right|.
\end{eqnarray}
This, indeed, explains why the steady states under RWA with losses are $\kappa$ and $g$ independent. By using Eq.~(\ref{ssgf}) or by studying the solution of the master equation~(\ref{slme}), for example by the method of solving first order differential equations via eigenvalues and eigenvectors or the pseudomode approach~\cite{mspsg}, we can determine the analytic form of the steady states for the extended Werner-like initial states~(\ref{initialstate}). For $\rho_{\Psi_{\alpha}^{\pm}}(0)$ type initial states, the steady states can be found as:
\begin{eqnarray}\label{sspsi}
\rho_{\Psi_{\alpha}^{\pm}}^{SS}=\left(\frac{3+r}{4}\right)\left|gg0\right\rangle\left\langle gg0\right|+\left(\frac{1-r}{4}\right)\left|\Phi^-0\right\rangle\left\langle\Phi^-0\right|.\nonumber\\
\end{eqnarray}
For $r=1$, i.e., for initially pure Bell-like states, $\left|\Psi_{\alpha}^{\pm}\right\rangle$,  $\rho_{\Psi_{\alpha}^{\pm}}(0)$ has no overlap with the subradiant state and all the excitations are lost in time evolution regardless of what $\alpha$ is, while for $0\leq r<1$, there would be a non-vanishing overlap term with the subradiant state determined by $r$ and the atom-field system will decay to a trapping state which is superposition of $\left|gg0\right\rangle$ and $\left|\Phi^-0\right\rangle$. On the other hand, for $\rho_{\Phi_{\alpha}^{\pm}}(0)$ type initial states, the steady states can be determined as:
\begin{eqnarray}\label{ssphi}
\rho_{\Phi_{\alpha}^{\pm}}^{SS}&=&\left(\frac{3-r}{4}\pm r\alpha\sqrt{1-\alpha^2}\right)\left|gg0\right\rangle\left\langle gg0\right|\nonumber\\
&+&\left(\frac{1+r}{4}\mp r\alpha\sqrt{1-\alpha^2}\right)\left|\Phi^-0\right\rangle\left\langle\Phi^-0\right|.
\end{eqnarray}
One should note  that $\rho_{\Phi_{\alpha}^{\pm}}(0)$ have always a non-vanishing term~[except for $\rho_{\Phi_{\alpha}^+}(0)$  at $\alpha=\frac{1}{\sqrt{2}}, r=1$] with the subradiant state, so the steady states are highly $\alpha$ and $r$ dependent.

$\left|\Phi^-\right\rangle$ is also the subradiant state for two-qubit $H_R$, which can be proven by noting that the Hilbert space $H_0$ of the Rabi model can be separated into two uncoupled subspaces $H_1$ and $H_2$ with $H_1=\frac{1}{\sqrt{2}}\left(\left|eg\right\rangle-\left|ge\right\rangle\right)\otimes H_{cavity}$ where the separation is also valid under decoherence mechanism and at the long time limit the state in $H_1$ becomes the dark state, $\frac{1}{\sqrt{2}}\left(\left|eg\right\rangle-\left|ge\right\rangle\right)\otimes\left|0\right\rangle$, independent of the coupling strength. Since $H_{JC}$ is the limiting case of $H_R$ and the master equation~(\ref{newme}) can describe the cavity losses in the strong-coupling regime, we conjecture that such a simple structure for the atom-field steady states as in Eqs.~(\ref{sspsi}) and~(\ref{ssphi}) should exist in the strong-coupling regime also. Below we will provide numerical evidence for our conjecture.
\begin{figure}[!hbt]\centering
{\scalebox{0.8}{\includegraphics{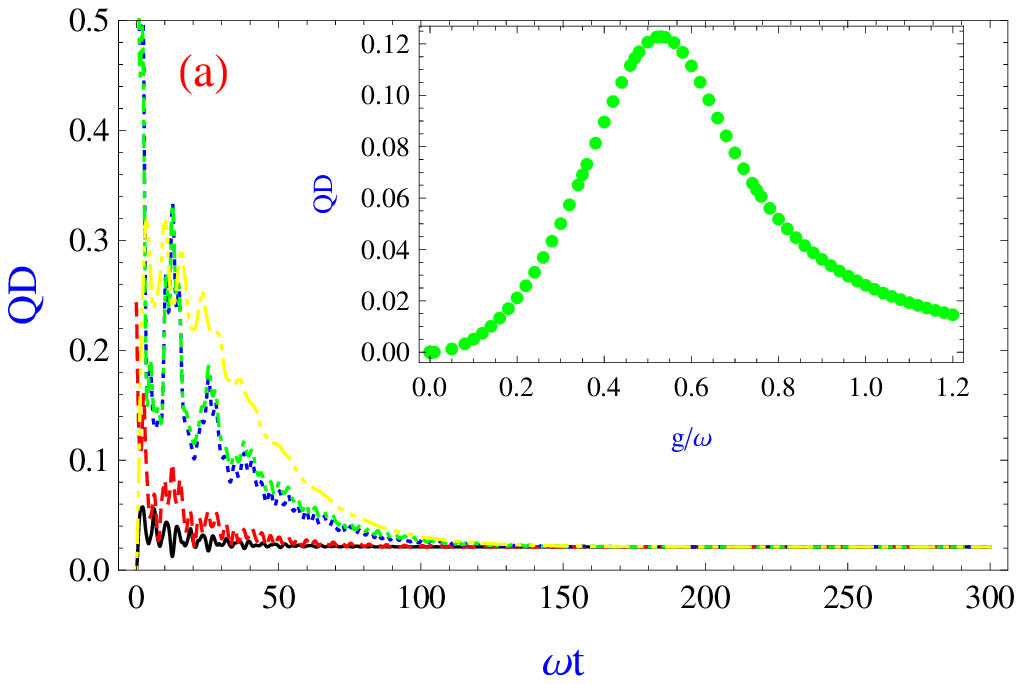}}}
{\scalebox{0.8}{\includegraphics{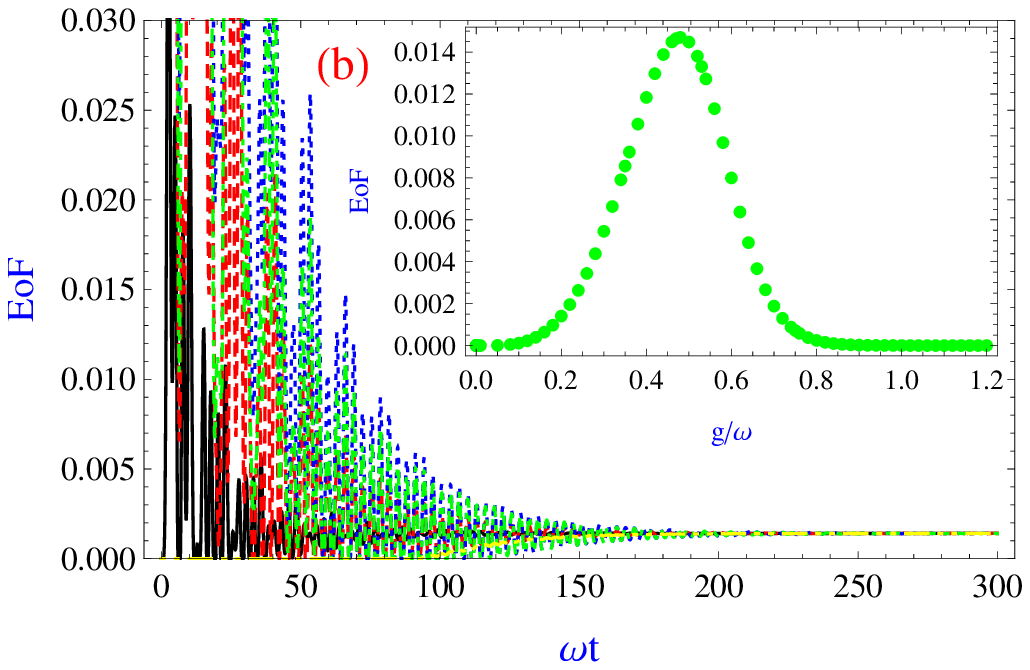}}}
\caption{(Color online) Effects of degree of correlations $\alpha$ on (a) QD and (b) EoF versus $\omega t$ for $\rho_{\Psi_{\alpha}^+}(0)$ initial state with $r=1$, $g=0.2\omega$, $\kappa=0.1\omega$, and $\alpha=0$~(black, solid), $\alpha=0.2$~(red, dashed), $\alpha=1/\sqrt{2}$~(blue, dotted), $\alpha=0.8$~(green, dot-dashed) and $\alpha=1$~(yellow, dot-dot-dashed). Here the results are obtained by using two-qubit $H_R$ and the master equation, Eq.~(\ref{newme}). Inset: QD and EoF of the ground state versus $g/\omega$. Around $g/\omega\approx 1.3$, the ground state of the Rabi model becomes almost twofold degenerate. So, we consider only $g/\omega\in [0,1.2]$ range in the insets.}
\end{figure}

We start to analyze the dynamics of correlations in the strong-coupling regime by using the improved master equation~(\ref{newme}) for the type of initial states which have no overlap with the subradiant state. Thus, in Fig.~2 we display QD and EoF versus $\omega t$ for the initial state $\rho_{\Psi_{\alpha}^+}(0)$ with $r=1$, $g=0.2\omega$, and $\kappa=0.1\omega$ and for several $\alpha$ values. In the case of RWA, the steady state is the vacuum which has no types of correlations. On the other hand, as shown in Fig.~2, the cavity decay in the strong-coupling regime drives the atom-field to a state which can carry appreciably high quantum discord~($\mathcal{C}_{QD}=0.021$) and entanglement~($\mathcal{C}_{EoF}=0.0014$) independent of what $\alpha$ is. In fact, we have observed that for the $\rho_{\Psi_{\alpha}^{\pm}}(0)$ type of initial states with $r=1$, the atom-field steady state is the ground state of two-qubit $H_R$ regardless of $g$. One can observe that the most important difference between the RWA and non-RWA dynamics at the long-time limit is the existence of non-zero QD and EoF created by the virtual processes; for RWA dynamics the necessary condition for the long-lived correlated state is the overlap of initial state with the subradiant state~\cite{sfzgp,mspsg,zfrt,fcah}, while this can be achieved without the overlap under non-RWA conditions because of the entangled nature of the ground state. In the insets of Fig.~2 we display the $g$ dependence of the ground state QD and EoF. Both QD and EoF as a function of $g/\omega$ display a peculiar structure; they have a single maximum at the intermediate values of $g/\omega$. Furthermore, $g/\omega$ dependence of EoF is almost Gaussian. For small $g/\omega$, where the RWA regime is valid, the QD and EoF are nearly zero, since the ground state is close to the vacuum state, while the correlations are high between $0.2<g/\omega<0.7$. Note that for the $g/\omega$ regions, where EoF is absent, non-zero QD still exists especially for $g/\omega>0.8$. Also note that the ground state QD and EoF show peaks at different $g/\omega$ values; QD is maximum nearly at $g/\omega\approx 0.53$, while EoF is at $g/\omega\approx 0.48$. On the other hand, the decrease in both QD and EoF as a function of $g/\omega$ at very strong-coupling regime might be an indication of the inability of the Rabi model to account for the possible nonlinear effects at large $g$ values, which may require higher order terms in the description of atom-field interactions~\cite{dzzzg}.
\begin{figure}[!hbt]\centering
{\scalebox{0.8}{\includegraphics{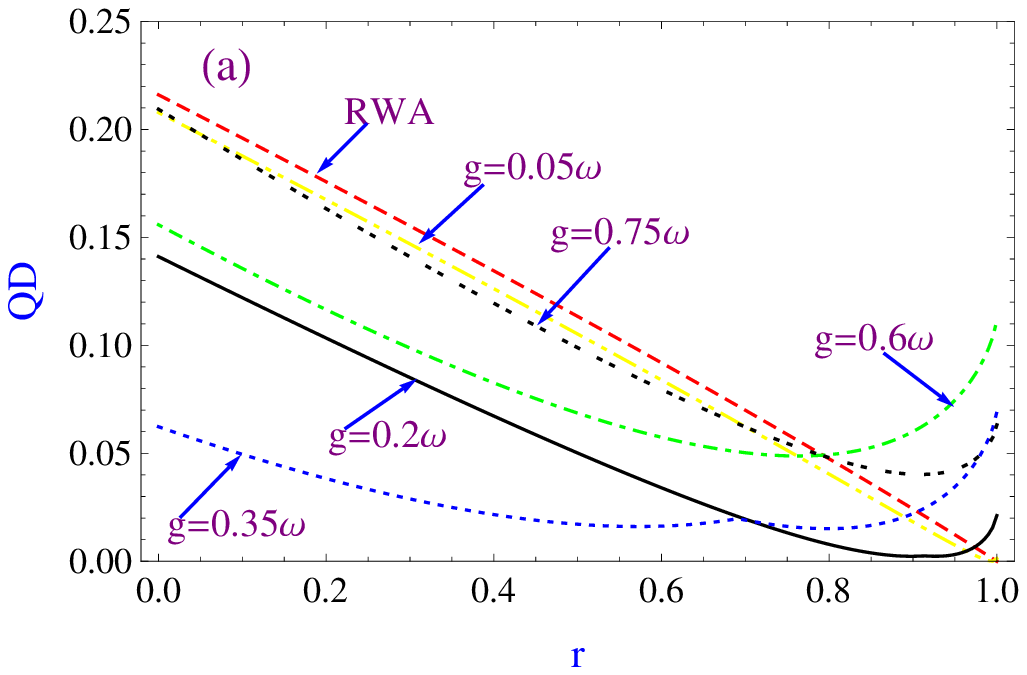}}}
{\scalebox{0.8}{\includegraphics{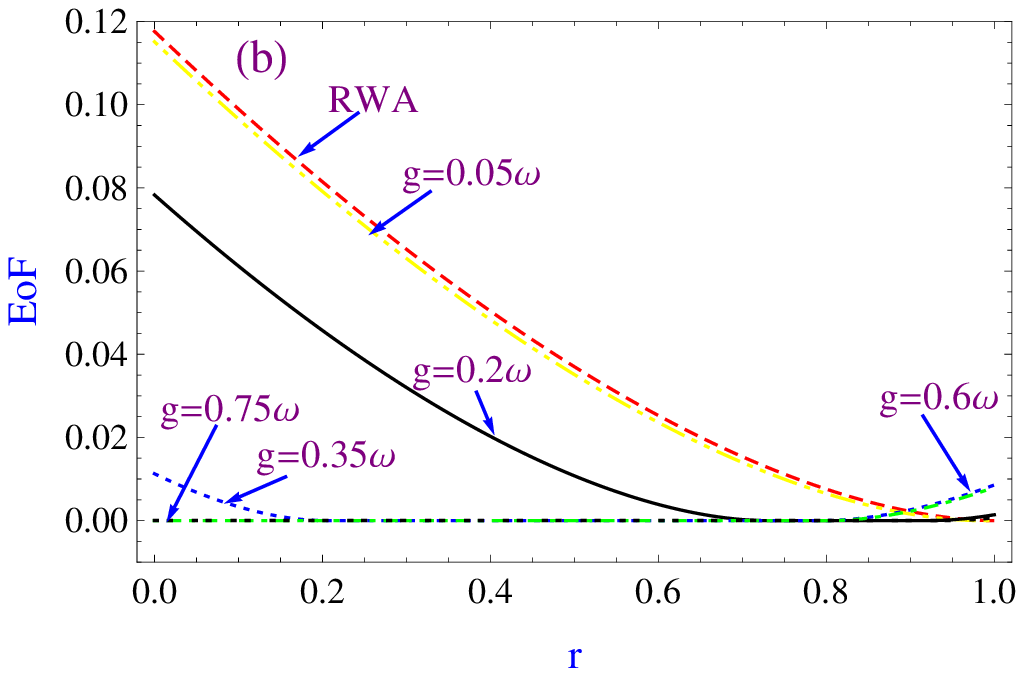}}}
\caption{(Color online) Steady state (a) QD and (b) EoF versus initial purity $r$ for $\rho_{\Psi_{\alpha}^{\pm}}(0)$ initial states under RWA~(red, dashed) and non-RWA dynamics with $g=0.05\omega$~(yellow, dot-dot-dashed),  $g=0.2\omega$~(black, solid), $g=0.35\omega$~(blue, dotted),  $g=0.6\omega$~(green, dot-dashed) and $g=0.75\omega$~(black, long-dotted). Note that for $g<0.01\omega$, the results obtained by the new master equation~(\ref{newme}) coincide with the RWA case~(red, dashed) given by Eq.~(\ref{sspsi}) and the steady states for both cases are $\alpha$ independent.}
\end{figure}

Special attention should also be given to the initial Bell $\left|\Phi^{+}\right\rangle=\frac{1}{\sqrt{2}}\left(\left|eg\right\rangle+\left|ge\right\rangle\right)$ and Bell-like $\left|\Psi_{\alpha}^{\pm}\right\rangle=\alpha\left|ee\right\rangle\pm\sqrt{1-\alpha^2}\left|gg\right\rangle$ states in the strong-coupling regime, since the entanglement dynamics have been extensively analyzed by using the standard dissipator~(\ref{slme}) under losses~\cite{fjl,jlmz,jlf,fare}. For $\kappa\geq 0.1\omega$, entanglement between atoms becomes zero or useless~(smaller than $10^{-4}$)  for a wide range of $g$, especially between $0.2<g/\omega<1$ under the standard dissipator case~\cite{fjl,jlmz,jlf,fare}. Contrary to those findings, one can see that such states can contain appreciably high EoF for $0.2<g/\omega<0.7$ independent of $\kappa$ as shown in the inset of Fig.~2(b), when one uses the improved master equation.

Now we consider the dynamics of initial states that have non-zero overlap with the dark state, $\left|\Phi^-0\right\rangle$. The effect of $r$  for $\rho_{\Psi_{\alpha}^{\pm}}(0)$ initial states on the dynamics of QD and EoF under losses have been investigated and the $r$ dependence of the steady state correlations are plotted in Fig.~3 for both RWA and non-RWA cases. Before discussing the role of counter-rotating terms on the steady state quantum correlations as a function of $r$, we should stress that a detailed analysis demonstrates that the atom-field steady states in the strong-coupling regime are in the form of Eq.~(\ref{sspsi}) with just $\left|gg0\right\rangle$ replaced by the ground state, $\left|\widetilde{gg0}\right\rangle$, of two-qubit $H_R$ and can be written as:
\begin{eqnarray}\label{ssnonpsi}
\rho_{\Psi_{\alpha}^{\pm}}^{SS}=\left(\frac{3+r}{4}\right)\left|\widetilde{gg0}\right\rangle\left\langle \widetilde{gg0}\right|+\left(\frac{1-r}{4}\right)\left|\Phi^-0\right\rangle\left\langle\Phi^-0\right|.\nonumber\\
\end{eqnarray}
Since the dark state of two-qubit $H_{JC}$ is also the dark state of two-qubit $H_R$, the form of the atom-field steady state for non-RWA Hamiltonian~[Eq.~(\ref{ssnonpsi})] is in expected form based on the same arguments that lead to Eq.~(\ref{sspsi}). More precisely, the overlap of the atomic initial state with the subradiant state is trapped in the subradiant state, while the remaining part decays to the ground state of the open quantum system under consideration. Also note that in the small $g$ limit $\left|\widetilde{gg0}\right\rangle$ reduces to $\left|gg0\right\rangle$. Therefore, the overall role of the counter-rotating term contributions in the steady states stems from the ground state of the two-qubit Rabi Hamiltonian. We have checked the form of Eq.~(\ref{ssnonpsi}) extensively by numerical means and found that it holds for all the initial states considered in the present work.

The initial purity dependence of steady state QD and EoF displayed in Fig.~3 for the RWA case shows an interesting feature; steady state values of both correlation measures show an inverse dependence on $r$. Since $r$ is a measure of initial mixedness of the state and the dissipation dynamics lead to an increase of the mixedness of the initial state, this finding might seem to be contradictory, but it stems from the fact that as $r$ decreases the overlap between the initial state and the maximally correlated dark state increases and the final steady state would contain a larger fraction of $\left|\Phi^-\right\rangle$ which leads to higher QD and EoF. In the strong-coupling regime, the ground state is $g$ dependent, which makes QD and EoF dependent on $g$ also. In fact, the counter-rotating contributions complicate the simple RWA picture and can induce separable atomic steady states for a wide range of $r$. Similar to the RWA case, the steady state correlations can display inverse $r$ dependence for a wide range, but contrary to the RWA case, the correlations can also be linearly $r$ dependent in the strong-coupling regime. Moreover, a generic result can be obtained from the comparison of the magnitudes of QD and EoF in Fig.~3; for the regions where EoF is zero (especially for strong couplings), appreciably high QD can still exist in the steady states.

To further elucidate the role of initial state and virtual processes on the steady state quantum correlations, we display QD versus $\alpha$ for $\rho_{\Phi_{\alpha}^{\pm}}(0)$ with $r=1$ in Fig.~4 and QD versus $r$ for the same initial states with $\alpha=0.2$ in the insets of Fig.~4. Similar to the $\rho_{\Psi_{\alpha}^{\pm}}(0)$ case, our detailed numerical analysis shows that the atom-field steady states under non-RWA dynamics for  $\rho_{\Phi_{\alpha}^{\pm}}(0)$ type initial states are in the form of Eq.~(\ref{ssphi}) with the vacuum state replaced by the ground state of the Rabi Hamiltonian, i.e., 
\begin{eqnarray}\label{ssnonphi}
\rho_{\Phi_{\alpha}^{\pm}}^{SS}&=&\left(\frac{3-r}{4}\pm r\alpha\sqrt{1-\alpha^2}\right)\left|\widetilde{gg0}\right\rangle\left\langle \widetilde{gg0}\right|\nonumber\\
&+&\left(\frac{1+r}{4}\mp r\alpha\sqrt{1-\alpha^2}\right)\left|\Phi^-0\right\rangle\left\langle\Phi^-0\right|.
\end{eqnarray}
Concisely analyzing the steady state QD dependence on the initial state parameters under RWA~(red, dashed line) signifies that the magnitude of QD directly depends on the amount of  overlap with the subradiant state. On the other hand, strong virtual processes can stir the simple RWA picture and interestingly can reverse the initial state parameter dependence of the steady QD compared to the RWA case. The $\alpha$ and $r$ dependence of steady state EoF is qualitatively similar to that of QD with a difference as previously indicated in Fig.~3(b); EoF can suffer death in the steady states for a wide range of $\alpha$ and $r$ under non-RWA dynamics, especially at $g=0.6\omega$ and $g=0.75\omega$. Thus, they are not plotted here.
\begin{figure}[!hbt]\centering
{\scalebox{0.8}{\includegraphics{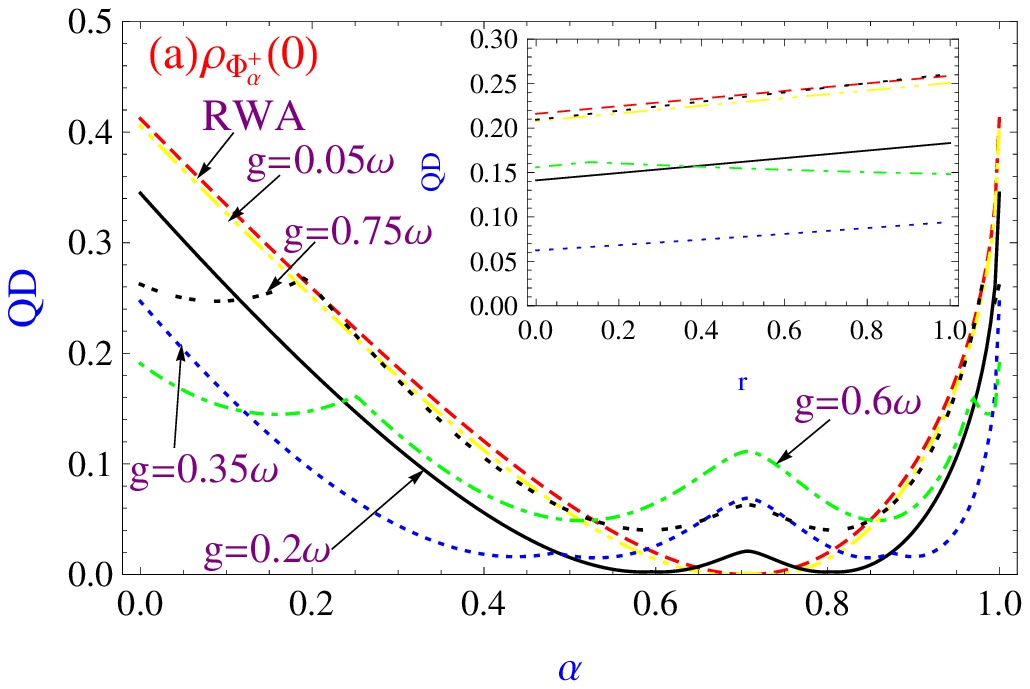}}}
{\scalebox{0.8}{\includegraphics{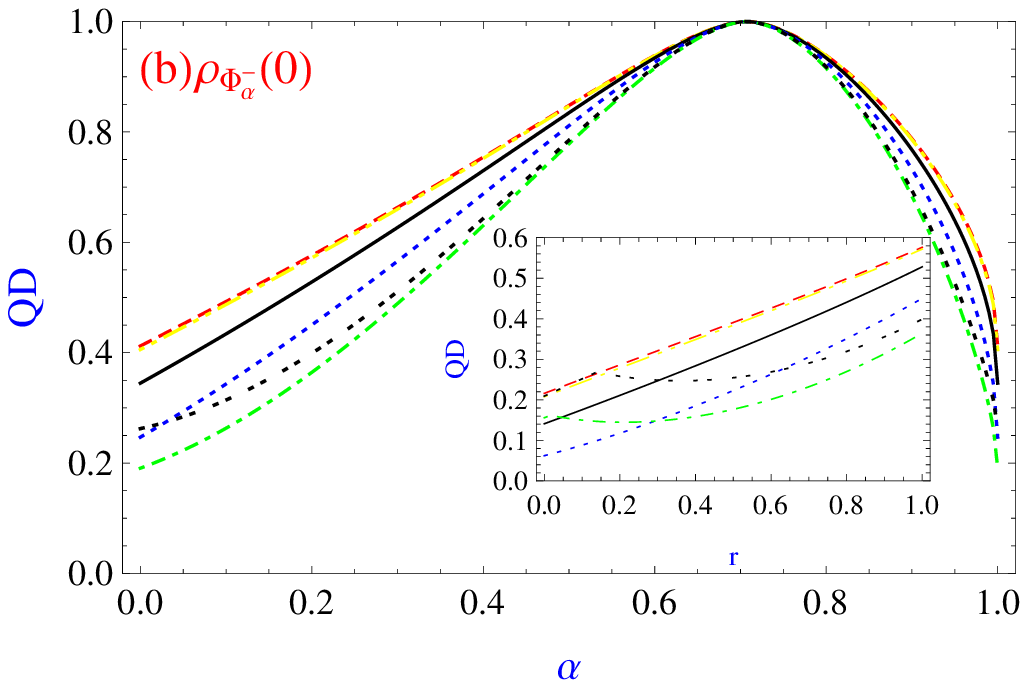}}}
\caption{(Color online) Steady state QD versus degree of correlations $\alpha$ for (a) $\rho_{\Phi_{\alpha}^{+}}(0)$ and (b) $\rho_{\Phi_{\alpha}^{-}}(0)$ initial states with $r=1$ under RWA~(red, dashed) and non-RWA dynamics with $g=0.05\omega$~(yellow, dot-dot-dashed),  $g=0.2\omega$~(black, solid), $g=0.35\omega$~(blue, dotted),  $g=0.6\omega$~(green, dot-dashed), and $g=0.75\omega$~(black, long-dotted). Inset: Steady state QD versus $r$ for (a) $\rho_{\Phi_{\alpha}^{+}}(0)$ and (b) $\rho_{\Phi_{\alpha}^{-}}(0)$ initial states with $\alpha=0.2$ and the same $g$ values used in the subfigures. Note that for $g<0.01\omega$, the results obtained by the new master equation~(\ref{newme}) coincide with the RWA case~(red, dashed) given by Eq.~(\ref{ssphi}).}
\end{figure}

\section{Conclusion}\label{conc}
We have studied a model of two cold-trapped two-level atoms strongly coupled to a single-mode leaky cavity field and initially prepared in extended Werner-like states. The dynamics of the atom-field system and quantum correlations between atoms, such as quantum discord and entanglement in the strong-coupling regime, have been investigated by using the recently derived Lindblad type master equation which offers quantum jumps among the eigenstates of the Hamiltonian of the open quantum system under consideration. Our results demonstrate that for extended Werner-like initial states, the amount of overlap between initial and dark states remains intact in the time evolution, while the other part decays to the ground state of the atom-field system. This result is found to hold for all atom-field coupling strengths and coincides at the weak-coupling regime with the results under RWA studied by the standard Lindblad master equation. The presence of counter-rotating terms are found to break down some of the recently predicted results for RWA dynamics, for example, they lead to quantum-correlated atomic steady states without the overlap of an initial state with the dark state which is crucial for RWA conditions. Moreover, the virtual processes can reverse the initial state parameter dependence of steady state QD and EoF contrary to the RWA case which is strongly  initial-dark state overlap dependent. However, we should stress here that the strong coupling regime seems to have quantum advantages compared to the RWA case only for initial states which lead to low~($\mathcal{C}_{QD}<0.05$) and zero discordant atomic steady states for the RWA case. Specifically, the strong virtual processes can induce separable atomic steady states.

The results obtained in the present study are quite simple and general. More importantly, the developed ideas can be expanded for local or global environment cases including multipartite and other decoherence channels, such as dephasing and spontaneous emissions of qubits and fields. For instance, for a model system with two qubits locally trapped  inside  leaky cavity fields~\cite{bellomo}, the dark state no longer exists. Therefore, it is expected that the atom-field steady states under cavity decay would always be the ground state of the system Hamiltonian. However, a more detailed discussion for this subject goes beyond the aim of the present study.


\section*{References}
\begin{thebibliography}{99}

\bibitem{rabi} I.I.~Rabi, Phys. Rev. {\bf 49}, 324 (1936).

\bibitem{ashhab} S.~Ashhab, F.~Nori, Phys. Rev. A {\bf 81}, 042311 (2010); S.~Ashhab, Phys. Rev. A {\bf 87}, 013826 (2013).

\bibitem{rbrev} For a brief  review see E.~Solano, Physics {\bf 4}, 68 (2011).

\bibitem{rbused} D.~Englund et al., Nature~(London) {\bf 450}, 857 (2007); T.~Niemczyk et al., Nature Phys. {\bf 6}, 772 (2010); A.T.~Sornborger, A.N.~Cleland, M.R.~Geller, Phys. Rev. A {\bf 70}, 052315 (2004); D.~Liebfried, R.~Blatt, C.~Monroe, D.~Wineland, Rev. Mod. Phys. {\bf 75}, 281 (2003); A.~Wallraff et al., Nature~(London) {\bf 431}, 162 (2004); P.F.~Diaz et al., Phys. Rev. Lett. {\bf 105}, 237001 (2010); A.~Federov et al., Phys. Rev. Lett. {\bf 105}, 060503 (2010); J.~Casanova et al., Phys. Rev. Lett. {\bf 105}, 263603 (2010).

\bibitem{rbsol} D.~Braak, Phys. Rev. Lett. {\bf 107}, 100401 (2011).

\bibitem{jcmodel} E.T.~Jaynes, F.W.~Cummings, Proc. IEEE {\bf 51}, 89 (1963).

\bibitem{burne} M. Brune, F.S.~Kaler, A.~Maali, J.~Dreyer, E.~Hagley, J.M.~Raimond, S.~Haroche , Phys. Rev. Lett. {\bf 76}, 1800 (1996).

\bibitem{jrkjlm} J.R.~Kuklinski, J.L.~Madajczyk, Phys. Rev. A {\bf 37}, 3175 (1988).

\bibitem{burne2} M.~Brune, S. Haroche, J. M. Raimond, L. Davidovich, N. Zagury,  Phys. Rev. A {\bf 45}, 5193 (1992).

\bibitem{wvhw} M.~Weidinger, B.T.H.~Varcoe, R.~Heerlein, H.~Walther, Phys. Rev. Lett. {\bf 82}, 3795 (1999).

\bibitem{grhw} G.~Rempe, H.~Walther, N.~Klein,  Phys. Rev. Lett. {\bf 58}, 353 (1987).

\bibitem{sppk} S.J.D.~Phoenix, P.L.~Knight, Phys. Rev. A {\bf 44}, 6023 (1991).

\bibitem{wddb} T.~Werlang, A.V.~Dodonov, E.I.~Duzzioni, C.J.V.~Boas, Phys. Rev. A {\bf 78}, 053805 (2008).

\bibitem{hnkb} H.T.~Ng, K.~Burnett, New J. Phys. {\bf 10}, 123014 (2008).

\bibitem{phoenix} S.J.D.~Phoenix, J. Mod. Opt. {\bf 38}, 695 (1991).

\bibitem{lemb} V.E.~Lembessis, Phys. Rev. A {\bf 78}, 043423 (2008).

\bibitem{bloch} F.~Bloch, A.~Siegert, Phys. Rev. {\bf 57}, 522 (1940).

\bibitem{mag} P.W.~Milonni, J.R.~Ackerhalt, H.W.~Galbraith, Phys. Rev. Lett. {\bf 50}, 966 (1983).

\bibitem{lindbadme} D.~Walls, G.~Milburn, {\it Quantum Optics} (Springer, Berlin, 1994).

\bibitem{bgb} F.~Beaudoin, J.M.~Gambetta, A.~Blais, Phys. Rev. A {\bf 84}, 043832 (2011).

\bibitem{slmeexp} S.~Haroche, J.-M.~Raimond, {\it Exploring the Quantum: Atoms, Cavities, and Photons} (Oxford University Press, Oxford, 2006).

\bibitem{sfzgp} S.~Maniscalco, F.~Francica, R.L.~Zaffino, N.L.~Gullo, F.~Plastina, Phys. Rev. Lett. {\bf 100}, 090503 (2008).

\bibitem{mspsg} L.~Mazzola, S.~Maniscalco, J.~Piilo, K.A.~Suominen, B.M.~Garraway, Phys. Rev. A {\bf 79}, 042302 (2009); L.~Mazzola, S.~Maniscalco, J.~Piilo, K.A.~Suominen, J. Phys. B: At. Mol. Opt. Phys. {\bf 43}, 085505 (2010).

\bibitem{zfrt} Z.~Ficek, R.~Tanas, Phys. Rev. A {\bf 77}, 054301 (2008).

\bibitem{fcah} C.H.~Fleming, N.I.~Cummings, C.~Anastopoulos, B.L.~Hu, J. Phys. A: Math. Theor. {\bf 45}, 065301 (2012).

\bibitem{fjl} Z.~Ficek, J.~Jing, Z.G.~Lu, Phys. Scr. {\bf T140}, 014005 (2010).

\bibitem{jlmz} J.~Jing, Z.G.~Lu, H.R.~Ma, H.~Zheng, J. Phys. B: At. Mol. Opt. Phys. {\bf 41}, 135502 (2008).

\bibitem{jlf} J.~Jing, Z.G.~Lu, Z.~Ficek, Phys. Rev. A {\bf 79}, 044305 (2009).

\bibitem{fare} F.~Altintas, R.~Eryigit, Phys. Lett. A {\bf 376}, 1791 (2012).

\bibitem{bellomo08} B.~Bellomo, R.~Lo~Franco, G.~Compagno, Phys. Rev. A {\bf 77}, 032342 (2008).

\bibitem{seke1} J.~Seke, Physica A {\bf 193}, 587 (1993).

\bibitem{seke2} J.~Seke, Quantum Opt. {\bf 4}, 151 (1992).

\bibitem{seke3} J.~Seke, Physica A {\bf 240}, 635 (1997).

\bibitem{brcpls} M.~Bina, G.~Romero, J.~Casanova, J.J.G.~Ripoll, A.~Lulli, E.~Solano, Eur. Phys. J. Special Topics {\bf 203}, 207 (2012).

\bibitem{rlsh} A.~Ridolfo, M.~Leib, S.~Savasta, M.J.~Hartmann, Phys. Rev. Lett. {\bf 109}, 193602 (2012).

\bibitem{entuse1} A.K.~Ekert, Phys. Rev. Lett. {\bf 67}, 661 (1991).

\bibitem{entuse2} C.H.~Bennett, G.~Brassard, C.~Grespeau, R.~Jozsa, A.~Peres, W.K.~Wotters, Phys. Rev. Lett. {\bf 70}, 1895 (1993).

\bibitem{wootters} W.K.~Wootters, Phys. Rev. Lett. {\bf 80}, 2245 (1998).

\bibitem{howz} H.~Ollivier, W.H.~Zurek, Phys. Rev. Lett. {\bf 88}, 017901 (2001). 

\bibitem{lhvv} L.~Henderson, V.~Vedral, J. Phys. A: Math. Gen. {\bf 34}, 6899 (2001). 

\bibitem{mbcpv} K.~Modi, A.~Brodutch, H.~Cable, T.~Paterek, V.~Vedral, Rev. Mod. Phys. {\bf 84}, 1655 (2012).

\bibitem{dqc1} E.~Knill, R.~Laflamme, Phys. Rev. Lett. {\bf 81}, 5672 (1998).

\bibitem{grover} J.~Cui, H.~Fan, J. Phys. A: Math. Theor. {\bf 43}, 045305 (2010).

\bibitem{rsp} B. Dakic, et al., arXiv:1203.1629.

\bibitem{wlnl} C.Z.~Wang, C.X.~Li, L.Y.~Nie, J.F.~Li, J. Phys. B: At. Mol. Opt. Phys. {\bf 44}, 015503 (2011).

\bibitem{facca} A.~Ferraro, L.~Aolita, D.~Cavalcanti, F.M.~Cucchietti, A.~Acin,  Phys. Rev. A {\bf 81}, 052318 (2010).

\bibitem{gds1} D.Z.~Rossato, T.~Werlang, E.I.~Duzzioni, C.J.V.~Boas, Phys. Rev. Lett. {\bf 107}, 153601 (2011).

\bibitem{gds2} T.~Werlang, S.~Souza, F.F.~Fanchini, C.J.V.~Boas, Phys. Rev. A {\bf 80}, 024103 (2009).

\bibitem{gds3} F.F.~Fanchini, T.~Werlang, C.A.~Brasil, L.G.E.~Arruda, A.O.~Caldeira, Phys. Rev. A {\bf 81} 052107 (2010).

\bibitem{gds4} B.~Wang, Z.Y.~Xu, Z.Q.~Chen, M.~Feng, Phys. Rev. A {\bf 81}, 014101 (2010).

\bibitem{gds5} L.~Mazzola, J.~Piilo, S.~Maniscalco, Phys. Rev. Lett. {\bf 104}, 200401 (2010).

\bibitem{gds6} F.~Ciccarello, V.~Giovannetti, Phys. Rev. A {\bf 85}, 010102 (2012).

\bibitem{gds7} A.~Streltsov, H.~Kampermann, D.~Brub, Phys. Rev. Lett. {\bf 107}, 170502 (2011).

\bibitem{gds8} J.~Maziero, L.C.~Celeri, R.M.~Serra, V.~Vedral, Phys. Rev. A {\bf 80}, 044102 (2009).

\bibitem{gds9} J.~Maziero, T.~Werlang, F.F.~Fanchini, L.C.~Celeri, R.M.~Serra, Phys. Rev. A {\bf 81}, 022116 (2010).

\bibitem{gds10} R.~Lo~Franco, B.~Bellomo, S.~Maniscalco, G.~Compagno,  Int. J. Mod. Phys. B {\bf 27}, 1345053 (2013).

\bibitem{gds11}  R.~Lo~Franco, B.~Bellomo, E.~Andersson, G.~Compagno, Phys. Rev. A {\bf 85}, 032318 (2012).

\bibitem{mswn} J.~Ma, Z.~Sun, X.~Wang, F.~Nori, Phys. Rev. A {\bf 85}, 062323 (2012). 

\bibitem{expss} For the most general initial states, one would also have coherence terms between the ground and subradiant states which leads to a very long-lived state rather than a stationary state~(see Ref.~\cite{fcah} for details). In the present study, the considered initial conditions do not have such coherence terms, i.e., $\left\langle gg\right|\rho_{AB}(0)\left|\Phi^-\right\rangle=0$.

\bibitem{dzzzg} L.H.~Du, X.F.~Zhou, Z.W.~Zhou, X.~Zhou, G.C.~Guo, Phys. Rev. A {\bf 86}, 014303 (2012). 

\bibitem{bellomo} B.~Bellomo, R.~Lo~Franco, G.~Compagno, Phys. Rev. Lett. {\bf 99}, 160502 (2007).


\end{thebibliography}
\end{document}